# Diffuse Interplanetary Radio Emission from a Polar Coronal Mass Ejection

Nat Gopalswamy*[(1)], Pertti Mäkelä[(2)], Seiji Yashiro[(2)], and Sachiko Akiyama[(2)]
(1) NASA Goddard Space Flight Center, Greenbelt, MD 20771, USA, https://cdaw.gsfc.nasa.gov
(2) The Catholic University of America, Washington DC 20064, USA

## Abstract

We report on the first detection of nonthermal radio emission associated with a polar coronal mass ejection. We call the radio emission as diffuse interplanetary radio emission (DIRE), which occurs in the decameter-hectometric wavelengths. The radio emission originates from the shock flanks that interact with nearby streamers.

## 1 Introduction

The starting frequency and the frequency extent of type II radio bursts are closely tied to the kinematics of the associated coronal mass ejections (CMEs) [1]. Type II bursts forming at or extending to frequencies below the ionospheric cutoff (~15 MHz) are associated with CMEs that are faster and wider on average [2]. One of the important characteristics of the solar sources of these bursts is that they are located in the active region belt, where the required high magnetic energy can be stored and released in the form of CMEs.

In this paper we report on nonthermal radio emission associated with a polar CME, whose source is located outside the active region belt. The radio burst is diffuse and appears like a noise storm but is closely related to the polar CME. On the other hand, the drift rate is similar to that of type II radio bursts. We call this diffuse interplanetary radio emission (DIRE) and show that the radio emission originates from the shock flanks.

## 2 Observations

The Radio and Plasma Wave (WAVES) experiment [3] on board Wind detected the DIRE. It consists of two radio receivers called Radio Receiver Band 1 (RAD1, 1040–20 kHz) and Band 2 (RAD2,13.825–1.075 MHz). The DIRE event is confined to the RAD2 band. The associated CME was observed by the Large Angle and Spectrometric Coronagraph (LASCO, [4]) on the Solar and Heliospheric Observatory (SOHO). The solar source was imaged by SOHO's Extreme-ultraviolet Imaging Telescope (EIT, [5]) and Yohkoh's Soft X-ray Telescope (SXT, [6]).

### 2.1 The DIRE

The Wind/WAVES radio dynamic spectrum in Fig. 1 (left) shows the DIRE in the 1-min resolution data. The DIRE starts at ~15:00 UT at the highest RAD2 frequency (~14 MHz). Initially, the DIRE shows FH structure, with the harmonic starting at ~15:30 UT at ~14 MHz. The 14-MHz H component is supposed to occur at ~7 MHz plasma level, indicating that the shock traveled ~30 min from the 14 MHz plasma level. Beyond about 16:00 UT, the FH structure is not observed and the DIRE becomes broadband. The lower edge of the DIRE appears to be the continuation of the F-component (see the plus symbols tracking the lower edge in Fig.1). The radio emission ends around 19 UT below 2 MHz. The appearance of DIRE with a series of bursts resembles a type III storm but with a narrower bandwidth (~3.5 MHz).

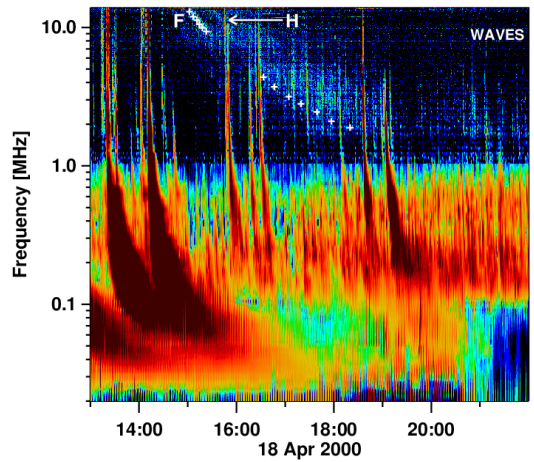

**Figure 1.** Wind/WAVES dynamic spectrum showing the diffuse interplanetary radio emission (DIRE) displayed with 1-min resolution. The plus symbols mark the fundamental (F) component starting at the upper end of the RAD2 receiver band at ~13.8 MHz. The harmonic (H) is weaker, but discernible in the 1-min dynamic spectrum. The type III bursts are unrelated to the DIRE.

The DIRE drift rate can be estimated from the FH structure at 14 MHz. Since the shock travels from 14 to 7 MHz plasma levels in ~30 min., the drift rate becomes 0.039 MHz s$^{-1}$. This is within the range of drift rates of type II bursts measured in the 1-14 MHz range [7-8]. Measuring the drift rates several times within the DIRE duration, we obtain a relation between the drift rate df/dt and emission frequency f as: df/dt = 6.89×10$^{-5}$f$^{1.58}$. The exponent 1.58 is similar to the one reported in [8] using data for a large number of type II bursts observed by different instruments. Thus, the DIRE resembles a normal type II burst in terms of drift rate, but very different in its morphology. The appearance is more like a herring-bone type II burst without the backbone that defines the type II burst. For DIRE, the drift is discerned from the overall envelope.

## 3 The Solar Source and CME Kinematics

Figure 2 shows two CMEs with the same first-appearance time (14:54:05 UT). The nose of the northern CME (CME-N) was heading along position angle (PA) 10º and originated from a compact active region (AR 8963) located at N17E12 and associated with a C4.2 GOES soft X-ray flare. The CME had a compact bright feature (~30º wide) with an extended shock-like structure that had a span of ~66º and extended mostly to right of the bright feature. When the CME first appeared in LASCO/C2 FOV, the leading edge was already at a height of 3.79 Rs. The post eruption arcade of CME-N is very compact as sown in Fig. 2c. The southern CME (CME-S) first appeared in the C2 FOV at 14:48 UT with a nose height of ~2.5 Rs at a PA of ~181º. CME-S was associated with the eruption of a polar crown filament. The PEA is very extended and occupies most of the southwest quadrant as can be seen from the Yohkoh/SXT image in Fig. 2c. The southern end of the PEA extends all the way to the south pole. The CME-S is very prominent with the familiar three-part structure with shock-like structure revealed by the adjoining streamers on both flanks that are affected by the shock as can be seen in: https://cdaw.gsfc.nasa.gov/movie/make_javamovie.php?date=20000418&img1=lasc2eit, which is a movie of the LASCO/C2 images. Although there are streamers on either side of CME-N in sky-plane projection, the movies do not show any indication that the CME affects them. This is because the narrow CME is from close to the disk center while the streamers are too far away from it to be affected.

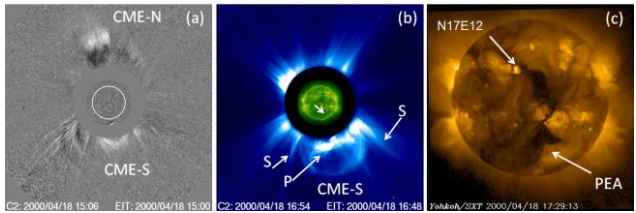

**Figure 2.** (a) SOHO/LASCO difference image at 15:06 UT showing two CMEs one from the north (CME-N) and the other from the south (CME-S). (b) CME-S in the direct image at 16:54 UT with the shock signatures (S) and the eruptive prominence (P). The location of eruption is pointed by an arrow in the SOHO/EIT image superposed on the LASCO/C2 image. (c) Sources of CME-N (N17E12) and CME-S (post-eruption arcade, PEA).

Figure 3 shows height-time (h-t) measurements with fitted curves: h = 3.80 + 1.16×$10^{-3}$t – 3.01×$10^{-8}$$t^2$ (CME-N) and h = 2.12 + 2.80×$10^{-4}$t + 2.16×$10^{-8}$$t^2$ (CME-S); t is in seconds from 14:54:05 UT and h is in Rs. CME-N slows down to ~340 km/s and CME-S speeds up to 530 km/s by 18:00 UT.

## 4 The DIRE Location Relative to the CME

In this section we show that CME-S is responsible for the DIRE and that the radio emission originates from the shock flanks where they interact with dense streamers based on the joint analysis of radio and white-light data.

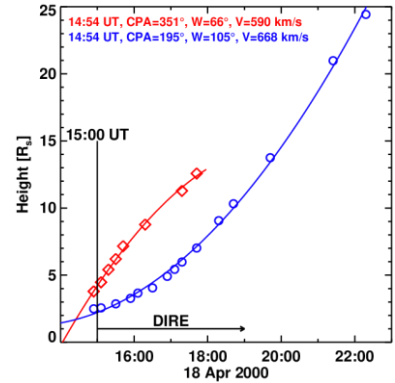

**Figure 3.** Height-time plots of CME-N (red) and CME-S (blue) showing decelerating and accelerating profiles. All measurements are in the sky plane and hence are subject to projection effects. The DIRE onset and duration are noted.

### 4.1 Electron Density in the Upstream Region

We use the LASCO/C2 polarized brightness image obtained before the eruption (2000 April 17 at 21:05 UT) to derive the electron density in the pre-event corona using the IDL routine pb_inverter. Valid electron densities are obtained only in certain position angles and heliocentric distances. Figure 4 shows the densities derived at various position angles at a heliocentric distance of 3 Rs. We use these densities to normalize a standard density model such as the one due to Leblanc et al. (1998) [9] to obtain radial density profiles at desired PAs. In other words, we force this model [9] to yield density values in Fig. 4 when r = 3 Rs. Figure 4 also shows that the sky-plane speed of the CMEs is >300 km/s when the DIRE is in progress. The deprojected speeds are expected to be high enough to drive a shock.

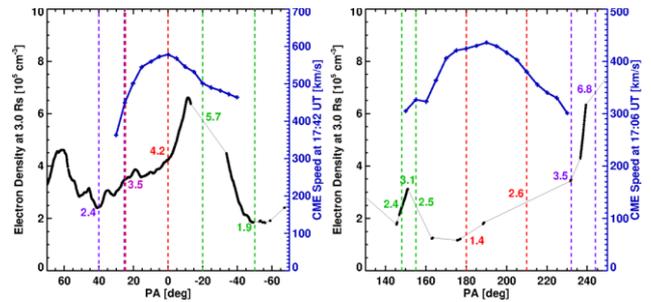

**Figure 4.** Pre-event coronal electron densities at 3 Rs as a function of PAs derived from polarization brightness (pB) images: CME-N (left) and CME-S (right). The CME speeds computed using second-order polynomial fit to h-t measurements made at 5º PA intervals are superposed. The pairs of vertical dashed lines mark the nose (red) and flank (green, blue) regions. The numbers indicate the electron densities at the boundaries of the nose and flank regions (in units of $10^5$ cm$^{-3}$). The thin lines show PA intervals where the density could not be determined from pB images.

Figure 5 shows the radial density profiles in the PA ranges of CME-N and CME-S. The horizontal black lines correspond to the plasma density at the onset (15:00 UT) and close to the end (18:00 UT) of the DIRE derived from the fundamental emission frequency. At the DIRE onset, the emission frequency is ~14 MHz, which corresponds to a local plasma density of 2.4×10⁶ cm⁻³. Such a density prevails around 2 Rs in the pre-event corona. The height-time plot in Fig. 3 shows that the nose of CME-N is already at ~4 Rs. It is unlikely that the flanks extend down to ~2 Rs. Towards the end of the DIRE event, around 18:00 UT, the emission frequency is ~2 MHz, so the local plasma density is ~5.3×10⁶ cm⁻³. Such a density prevails at a heliocentric distance of ~5 Rs. However, around this time, the CME-N nose is already at ~12.5 Rs and flanks are not discernible down to 5 Rs, where the DIRE needs to originate if CME-N is responsible. Therefore, we can rule out that CME-N is responsible for the DIRE emission.

In the case of CME-S, the DIRE source heights indicated by Fig. 5 (right) at the onset match with the CME heights (see Fig. 3 for the nose height at 15:00 UT). Close to the end of the DIRE event (~18:00 UT), the emission heights are in the range 4-5 Rs. Fig.3 shows that the nose is around 7 Rs. The LASCO images taken at 17:42 and 18:18 UT do show the flanks in the vicinity of ~5 Rs. Therefore, we conclude that CME-S responsible for the DIRE event.

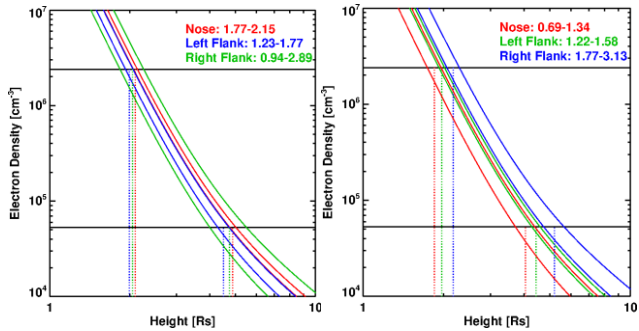

**Figure 5.** Radial profiles of electron density at the nose and flanks of CME-N (left) and CME-S (right). The range of multipliers to be used with the Leblanc et al. (1998) radial density profile so that it matches the observed densities at 3 Rs are indicated on the plots. The horizontal black lines indicate the plasma density in the DIRE source region at 15:00 UT (upper, 2.4×10⁶ cm⁻³) and 18:00 UT (lower, 5.3×10⁴ cm⁻³) obtained from the emission frequency (fundamental). The vertical dashed lines indicate the expected heights from which the DIRE originates.

### 4.2 Plasma Frequency vs. Position Angle

Since the DIRE has FH structure, we convert the electron density profile at various heights (see Fig. 5) into plasma frequency profiles as follows. (1) From h-t data obtain the CME leading-edge heights at every 5⁰ PA for selected times, linearly interpolating from observation times. (2) From the leading-edge heights and the density multipliers shown in Fig. 5, obtain the upstream electron density at 5⁰ PA intervals. (3) Convert the electron density to plasma frequency as plotted in Fig. 6 at various times from the onset to the end of the DIRE event. The horizontal lines indicate the DIRE fundamental emission frequency plotted on the plasma frequency curves in the PA range of CME-N and CME-S. We expect the DIRE emission to originate from the PA at which the DIRE line intersects the plasma frequency curve.

Considering CME-N in Fig. 6, we see that during the first hour of the DIRE emission, the horizontal line does not intersect the plasma curve. That means, the onset of the DIRE emission cannot be explained by CME-N, consistent with the discussion above. Beyond 16:00 UT, we see that emission frequency matches the plasma curve at PAs >60⁰. Since the central position angle (CPA) of CME-N is 351⁰, this implies a CME width >100⁰, which is larger than the observed width of CME-N. Furthermore, the speed at PA=60⁰ falls to 200 km/s and to even lower values beyond that. Thus, we confirm that CME-N is not responsible for the DIRE event.

In the case of CME-S, all the DIRE lines intersect with the plasma lines from the onset to the end of the DIRE event at the right flank (at PA >225⁰). On the left flank, the DIRE line at the onset does not intersect the 15:00 UT plasma line. After that the DIRE lines intersect both the right and left flanks. This suggests that the observed DIRE has contributions from both the flanks beyond ~16:00 UT. This may be the reason we see the fundamental and harmonic are separated in the beginning (contribution comes only from the right flank over a narrow section of the shock). After 16:00 UT, the fundamental and harmonic merge because of contributions coming from different sections of the shock, making the emission broadband.

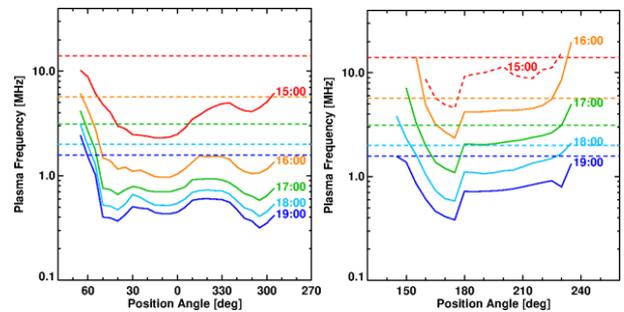

**Figure 6.** Plasma frequencies as a function of the PA centered on CME-N (left) and CME-S (right) at various times during the DIRE event. The horizontal dashed lines represent the DIRE frequencies at the fundamental. DIRE originates where a horizontal line intersects the plasma frequency curve of the same color.

The electron densities and plasma frequencies are integrated along the line of sight. The streamers and CMEs are sky-plane projections. Fortunately, the CME is far to the south so, a simple deprojection of CME height and speed should be sufficient. E.g., when the CME leading edge is at ~2 Rs, the deprojected height is ~2.6 Rs. Therefore, the densities derived from Fig. 5 for the nose and flanks should not be too different from reality.

## 5 Discussion

The primary finding of this paper is that a polar CME is capable of driving a shock and accelerating nonthermal electrons to produce the DIRE Event. Although the DIRE event appears like a type III storm, the drift rate is similar to that of normal type II radio bursts. Furthermore, the bandwidth of the DIRE event is much smaller than that of a typical type III storm. Type III storms in the DH domain are thought to be a continuation of type I storms at higher frequencies [10]. Type III storms have thousands of short-lived type III-like bursts in rapid succession [11] and the storm typically can last for days to weeks. The type III storms are not associated with eruptions and are thought to be due to small-scale, quasi-continuous energy releases in active regions.

The DIRE event is specifically tied to an eruption and the emission is identified with the shock flanks that interact with adjacent dense streamers. Therefore, we suggest that the DIRE is a variant of type II burst that originates due to CME-shock interaction with streamers. Liu et al. [12] noted the presence of type III storms in association with some intense type II bursts. They also noted that the "type III storm seems to have an overall drift rate similar to that of the type II burst". The difference between our event and those in [12] is that the latter were associated with intense type II bursts. We think their "normal" type II bursts come from the shock nose, while the "type III storm" comes from the shock flanks. In this sense, what they call as type III storm is indeed a DIRE event. We are confident about this because the type III storm was invariably found at higher frequencies than the associated type II burst. This can happen when the shock is curved with the nose at a larger height (lower plasma frequency) than the flanks (higher plasma frequency).

The interaction of the shock flanks with the streamer is an important aspect of the DIRE event. This can be understood from the fact that the expected scale height of the ambient density seems to be much shorter than what is expected for a normal corona. Recall that drift rate df/dt of the fundamental around f=14 MHz is 0.039 MHz s$^{-1}$. The shock speed is given by V = 2L(1/f)(df/dt), where L=r/α is the density scale height for a density N(r) ~ r$^{-α}$. For the RAD 2 frequency range, α ~4 [8]. Since r=2, we get L = 0.5 Rs and V = 1940 km/s at 15:00 UT. This speed is several times larger than what is indicated by the height-time measurements in the flank region. If we take the deprojected flank speed to be ~400 km/s, we get L = 0.1 Rs, which is smaller than the density scale height of the ambient corona by a factor of 5. Such short scale heights can result if the shock traverses streamer stalks that have a size scale of 0.1 Rs. The shock strength can also vary rapidly depending on the streamer density and magnetic field. While the flank region has low Alfven speed due to the presence of the streamer, the nose region has much higher Alfven speed because the ambient medium is tenuous in the polar region, as can be discerned from the direct image in Fig. 2b. Therefore, there is no type II burst from the nose region.

This is the first time a high-latitude CME has been found to be shock-driving. This provides additional evidence that polar CMEs [13] are similar to low-latitude CMEs from outside active regions [14], but generally of lower energy. In the present case, the energy is high enough to drive a shock and accelerate electrons that resulted in the nonthermal radio emission. The DIRE event highlights possible variability in the acceleration of nonthermal particles by shocks involving shock geometry and the inhomogeneities in the ambient medium [15].

# 6 Acknowledgements

We benefited from the open data policy of *SOHO*, *STEREO*, SDO, *GOES*, and *Wind* teams. Work supported by NASA's LWS and Heliophysics GI programs.